\documentclass[epj-spec]{svjour}
\usepackage{graphics}
\def\prl{{\em Phys. Rev. Lett. }}
\def\prc{{\em Phys. Rev. {\bf C} }}

\def\npa{{\em Nucl. Phys. {\bf A}}}

\def\epjc{{\em Eur. Phys. J. {\bf C}}}
\def\epja{{\em Eur. Phys. J. {\bf A}}}
\def\plb{{\em Phys. Lett. {\bf B}}}

\def\zpc{{\em Z. Phys. {\bf C}}}
\def\zpa{{\em Z. Phys. {\bf A}}}

\begin{document}
\title{Saturation of Transverse 
Energy per Charged Hadron and Freeze-Out Criteria in Heavy-Ion Collisions}
\author{J.~Cleymans\inst{1}\fnmsep\thanks{\email{Jean.Cleymans@uct.ac.za}} \and R.~Sahoo\inst{2} \and D.K.~Srivastava\inst{3} \and S.~Wheaton\inst{1}}
\institute{UCT-CERN Research Centre and Department  of  Physics,
University of Cape Town, Rondebosch 7701, South Africa \and Institute of Physics, Sachivalaya Marg, Bhubaneswar 751005, India \and Variable Energy Cyclotron Centre, 1/AF Bidhan Nagar, Kolkata 700064, India}
\abstract{
For beam energies from SPS to RHIC, 
the  transverse energy per charged particle, $E_T/N_{\textrm{ch}}$,
saturates at a value of approximately 0.8 GeV. 
A direct connection between  this value and 
the freeze-out criterium $E/N \approx 1$ GeV  for the primordial energy
and particle number in the hadronic resonance gas model is established.
} 
%
\maketitle
\begin{figure}
\begin{center}
\resizebox{0.75\columnwidth}{!}{%
\includegraphics{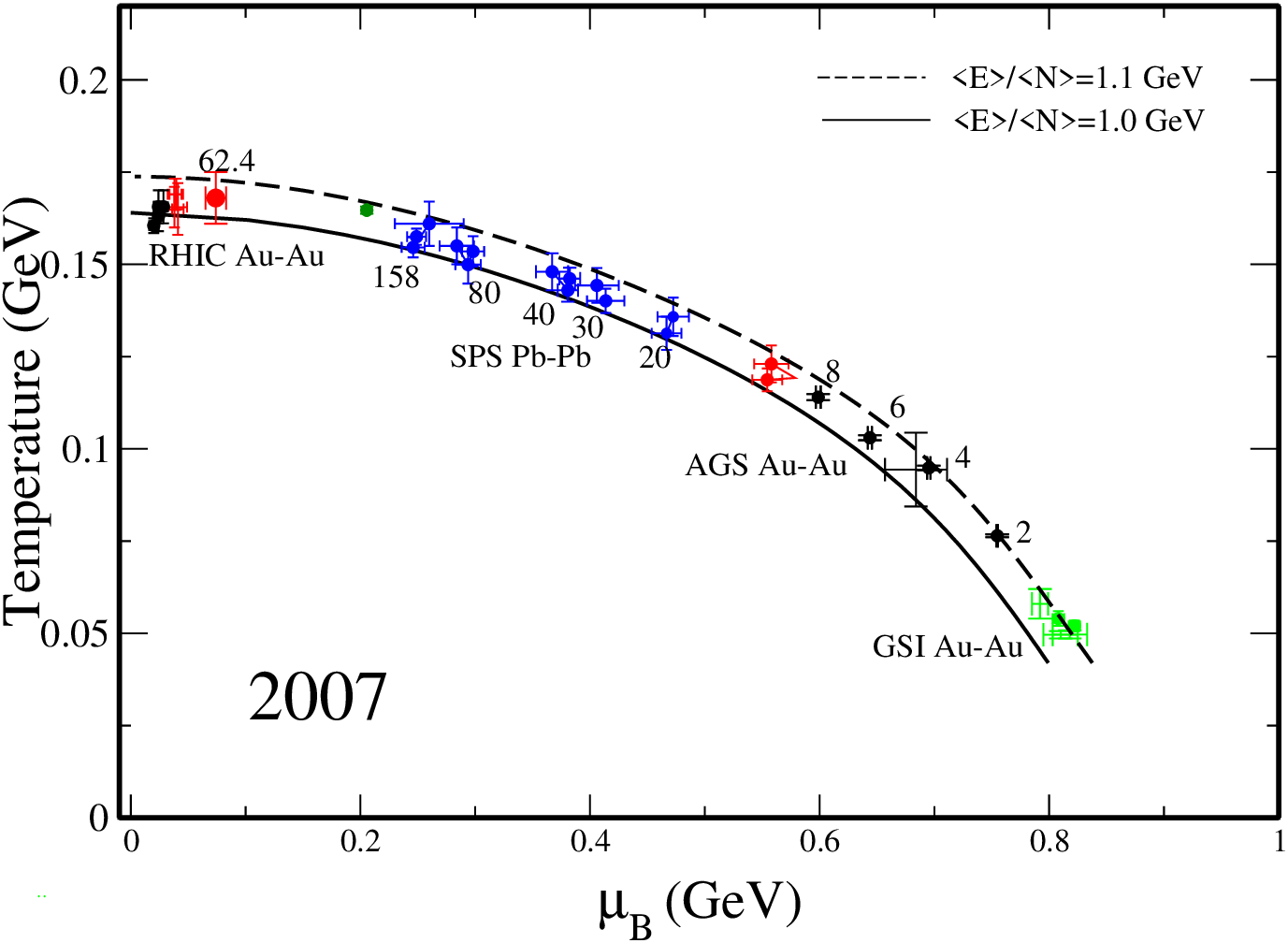}}
\caption{Temperatures and baryon chemical potentials 
deduced from yields in relativistic heavy-ion collisions 
at various beam energies. The lower AGS points, denoted 2, 4, 6 and 8 still have to 
be confirmed. The solid lines correspond to constant values of $E/N$ being kept fixed at 1 and 1.1 GeV
respectively. The point at 62.4 GeV has been taken from Ref.~\cite{takahashi}. 
}
\label{equilibrium}
\end{center}
\end{figure}

%
All relativistic heavy-ion experiments  have so far confirmed 
the 
validity of $E/N \approx$ 1 GeV as a freeze-out criterium,  with 
 $E$ and $N$ being, 
respectively the total energy and  particle number of the 
primordial hadronic resonances before they decay into stable hadrons,
 i.e the energy $E$ refers to the energy
of all hadronic resonances like $\rho, \Delta, \omega, \dots$ and the 
number $N$ refers to the total number of these particles at the chemical 
freeze-out point.
These quantities can 
not be determined directly from experiment unless the 
final state multiplicity is  low and hadronic resonances can be identified,
which is not the case in  relativistic heavy-ion collisions.
It is thus
not straightforward to link $E/N$ to directly measurable quantities. 
 In this paper we establish an approximate connection between $E/N$ and the 
ratio of the pseudo-rapidity density of transverse energy and that of
the charged particle yield, ($dE_T/d\eta/dN_{\textrm{ch}}/d\eta 
\equiv E_T/N_{\textrm{ch}}$),
for  beam energies ranging from about 1 AGeV up to 200 AGeV. 
In this energy range,
$E_T/N_{\textrm{ch}}$ at first increases rapidly from SIS~\cite{fopi}
  to AGS~\cite{e802,e814_6GeV}, then saturates to 
a value of about 800 MeV at SPS~\cite{wa98_17GeV,na49_17GeV,na49} energies 
and remains constant
up to the highest available RHIC energies~\cite{star200GeV,phenix130GeV,phenixSyst}. 
The present analysis of $E_T/N_{\textrm{ch}}$ 
uses the hadron resonance gas model (thermal model) which describes 
the final state in relativistic heavy-ion collisions as composed of 
hadrons, including heavy hadronic resonances
as being  in thermal and chemical equilibrium.
Our analysis therefore starts by relating the number of charged particles 
seen in the detector to the number of primordial hadronic resonances and 
the transverse energy to the energy $E$ of primordial hadrons.
The present status of $E/N$ is shown in Fig.~\ref{equilibrium}.
The same results can also be plotted differently using energy density and
baryon density as variables instead of $T$ and $\mu_B$. This brings out
very clearly the maximum in the baryon density. At higher energies the baryon 
density goes to zero due to the vanishing of $\mu_B$ \cite{randrup_prc}. 

\begin{figure}
\begin{center}
\resizebox{0.75\columnwidth}{!}{%
\includegraphics{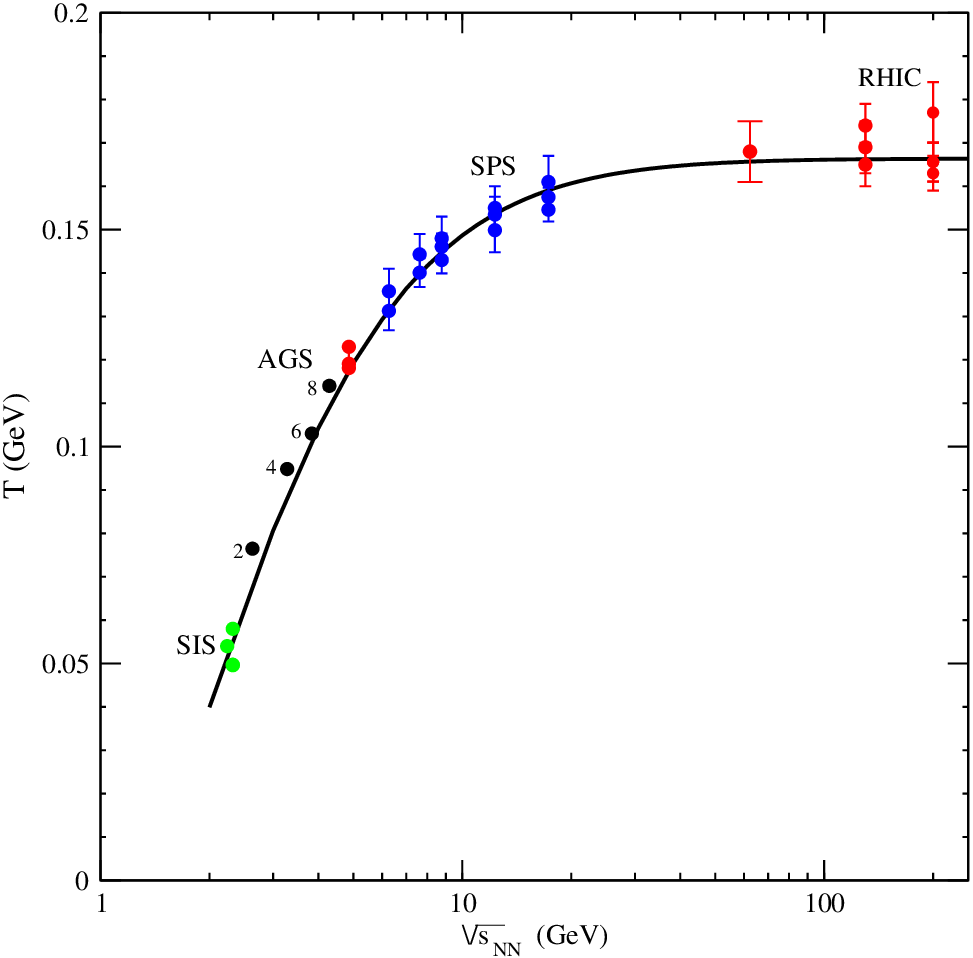}}
\caption{Saturation of the chemical freeze-out temperature at high energies.}
\label{T_e}
\end{center}
\end{figure}

\begin{figure}
\begin{center}
\resizebox{0.75\columnwidth}{!}{%
\includegraphics{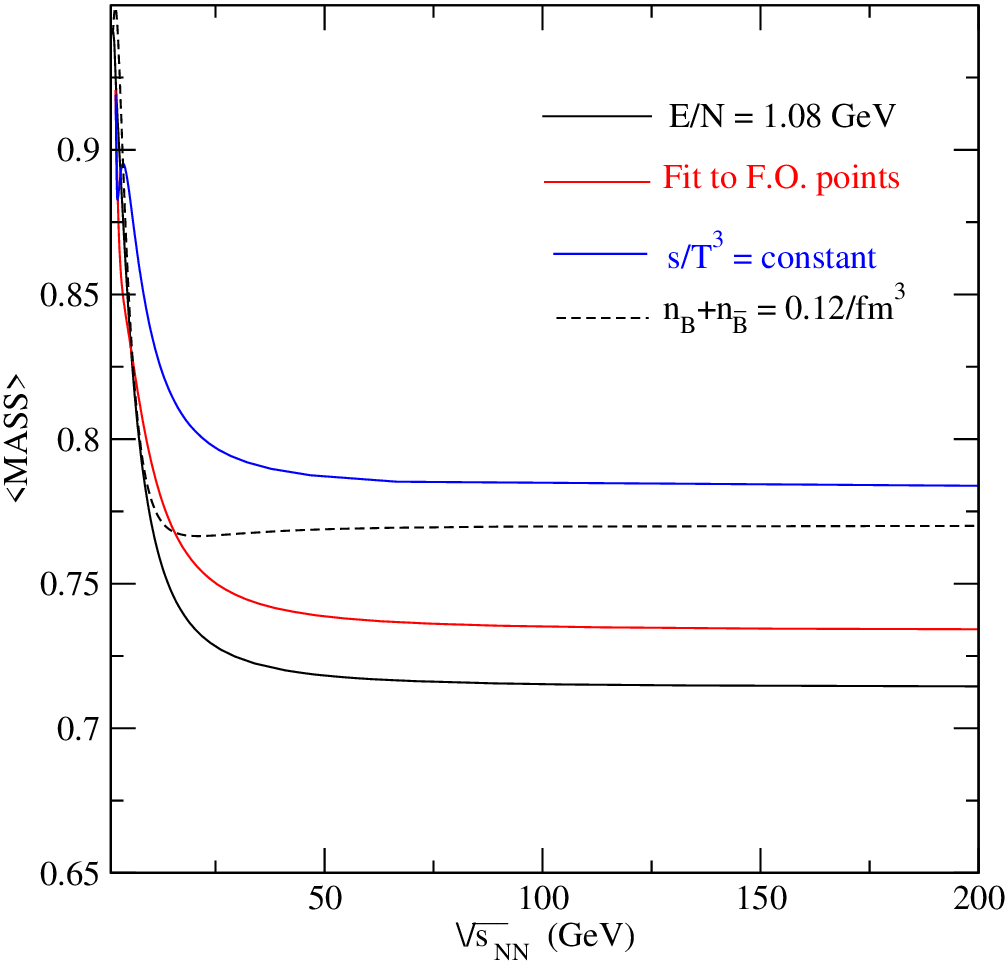}}
\caption{Saturation of the average mass in the hadronic resonance 
gas model at high beam energies for various freeze-out criteria 
proposed in the literature~\cite{PBMS,Majiec,TawfikEntropy,tawfik2}.
}
\label{meanmass}
\end{center}
\end{figure}

\begin{figure}
\begin{center}
\resizebox{0.75\columnwidth}{!}{%
\includegraphics{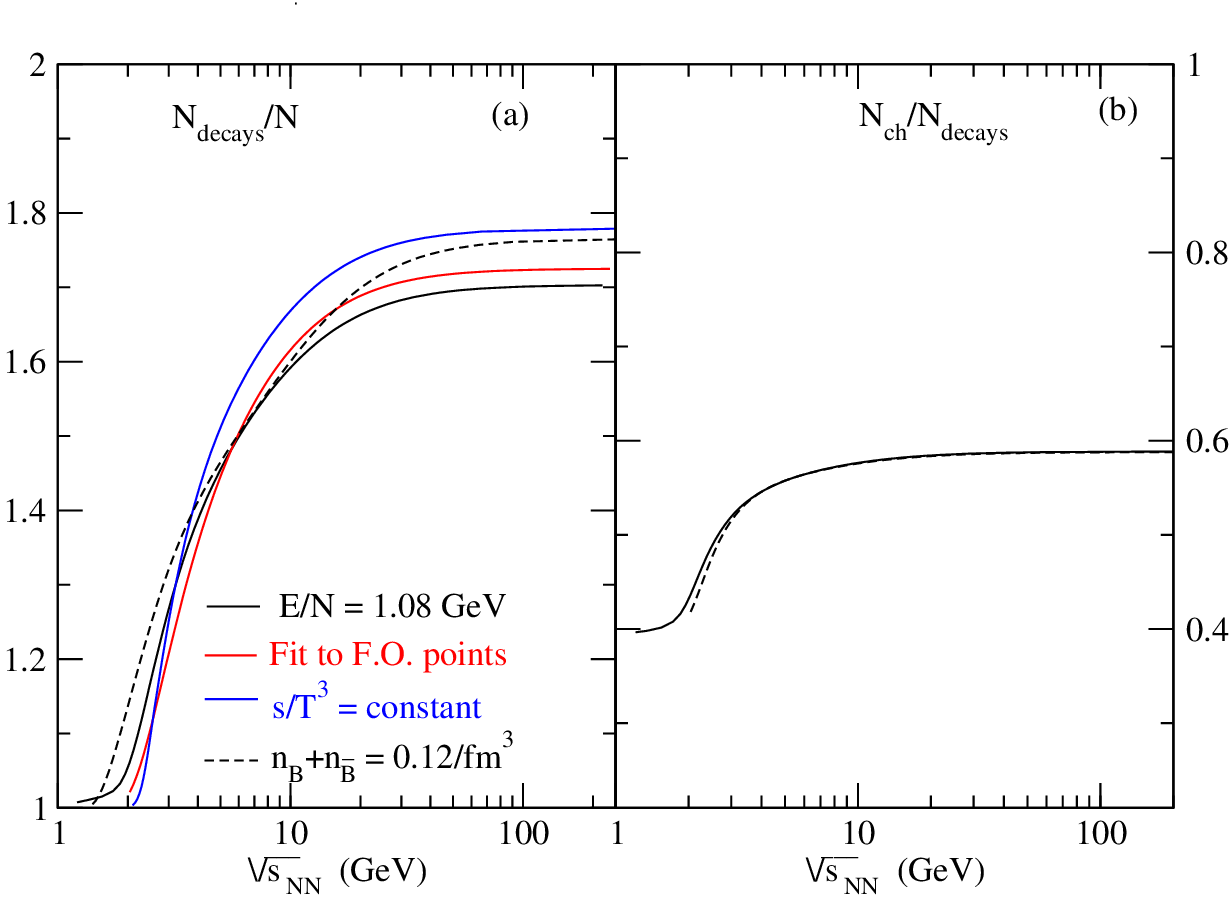}}
\caption{Saturation of $N_{\textrm{decays}}/N$ (a) and 
$N_{\textrm{ch}}/N_{\textrm{decays}}$ (b) 
with $\sqrt{s_{NN}}$. 
In (a) the results from various freeze-out criteria are indicated.
In (b) the different freeze-out criteria give
results that are indistinguishable.}
\label{joint1}
\end{center}
\end{figure}

In this paper all thermal model calculations were performed using the 
THERMUS package~\cite{THERMUS}. 
%
%
%
At high energies the chemical freeze-out temperature saturates at
a value of about 160 - 170 MeV as shown in Fig.~\ref{T_e} and at the same time
the baryon chemical potential becomes very small~\cite{FreezeOuts}.
As a consequence, several other quantities also become
independent of beam energy. The average mass of hadronic resonances
saturates at approximately the $\rho$ mass at high energies as shown
 in Fig.~\ref{meanmass}. The ratio of all hadrons after resonance decays
to the number of directly emitted hadrons at chemical freeze-out
saturates at a value of 
about 1.7 as shown in Fig.~\ref{joint1}. All of these are direct 
consequences of the saturation of the freeze-out temperature observed in 
Fig.~\ref{T_e} for increasing beam energies and the associated convergence 
of the baryon chemical potential to zero. 

The transverse energy, $dE_T/d\eta$,  is 
defined as the energy deposited transverse to the beam
direction in a given interval of pseudo-rapidity $\eta$, since this quantity 
is integrated over usually, we will write $E_T$ for brevity even though it has 
not been integrated over the full pseudo-rapidity interval.
The transverse energy  has two components, the hadronic one,
$E_T^{\textrm{had}}$,  and the electromagnetic one, $E_T^{\textrm{em}}$,
coming from the electromagnetic particles (photons, electrons and positrons).
Electromagnetic calorimeters are used to measure $E_T^{\textrm{em}}$ 
whereas hadronic
calorimeters or the Time Projection Chamber (for particle identification
and momentum information) are used to measure $E_T^{\textrm{had}}$.
The energy of a particle is defined as being the kinetic energy for nucleons, 
for anti-nucleons 
as the total energy plus the rest mass and for all other particles as the total
energy~\cite{star200GeV,phenix130GeV,helios}. 

In the experiments, 
the transverse energy and the 
charged particle multiplicity are measured in a similar way so that most of the systematic 
uncertainties cancel out in the ratio. 
Experiments  
have reported a constant value of the ratio  
$E_T/N_{\textrm{ch}}~\sim~0.8$ GeV from 
SPS to RHIC~\cite{star200GeV,phenixSyst}, 
with the ratio being almost independent of centrality of the 
collision for all measurements at different energies. 
In all cases the value of $E_T/N_{\textrm{ch}}$ has been taken for the 
most central collisions at mid-rapidity. At the end of this paper we consider 
the centrality dependence of  $E_T/N_{\textrm{ch}}$. 
When this ratio is observed for the full range of center of mass energies, 
it shows two regions~\cite{phenixSyst}. 
In the first region from lowest $\sqrt{s_{NN}}$ 
to SPS energy, there is a steep increase of the $E_T/N_{\textrm{ch}}$ ratio with $\sqrt{s_{NN}}$. 
In this regime, the increase of $\sqrt{s_{NN}}$ causes an increase in the $\left<m_T\right>$
 of the 
produced particles. In the second region,  SPS  to higher 
energies, the $E_T/N_{\textrm{ch}}$ ratio is very weakly dependent on 
$\sqrt{s_{NN}}$. 

To estimate  $E_T/N_{\textrm{ch}}$ in the thermal model 
we relate the number of charged particles,  $N_{\textrm{ch}}$,
to the number, $N$, of primordial hadrons.
To estimate the charged particle multiplicity at different center of mass energies from 
the thermal model, we proceed as follows. First we study the variation of the
ratio of the total particle multiplicity in the final state, $N_{\textrm{decays}}$, 
and that in the primordial
i.e. $N_{\textrm{decays}}/N$ 
with $\sqrt{s_{NN}}$.
This ratio starts from one, since there are only very few 
resonances produced at low beam energy 
and becomes almost independent of energy after SPS energy.
The value of $N_{\textrm{decays}}/N$ in the region where it is 
independent of $\sqrt{s_{NN}}$
is around 1.7. The excitation function of 
$N_{\textrm{decays}}/N$
is shown in Fig.~\ref{joint1}(a). Secondly, we have studied the variation of the ratio of
charge particle multiplicity and the particle multiplicity in the final state
($N_{\textrm{ch}}/N_{\textrm{decays}}$) with $\sqrt{s_{NN}}$. 
This is shown in Fig.~\ref{joint1}(b). 
The $N_{\textrm{ch}}/N_{\textrm{decays}}$ ratio starts around 0.4 at
lower $\sqrt{s_{NN}}$ and shows an energy independence at SPS and higher energies.
At lower SIS energy, the baryon dominance at mid-rapidity makes 
$N_{\textrm{ch}}/N_{\textrm{decays}}\sim N_{\textrm{proton}}/N_{\textrm{(proton+neutron)}}$ which 
has a value of 0.45 for  Au-Au collisions

As the next step we connect the transverse energy $E_T$ to the the energy of the primordial
hadrons $E$. 
In the hadronic resonance gas model there is a sum over 
all hadrons; furthermore, taking
into account the experimental configuration which leads to adding 
the mass of the nucleon for anti-nucleons and subtracting the same for
nucleons one has
\begin{eqnarray}
\left<E_T\right> & \equiv & V\sum_{i={\textrm{Nucleons}}}\int\frac{d^3p}{(2\pi)^3} (E_i-m_N)\sin\theta~~f(E_i) \nonumber\\
&&+V\sum_{i=\textrm{Anti-nucleons}}\int 
\frac{d^3p}{(2\pi)^3} (E_i+m_N) \sin \theta~~f(E_i)\nonumber\\
&&+V\sum_{i=\textrm{All ~Others}}\int \frac{d^3p}{(2\pi)^3} E_i \sin \theta~~f(E_i) , \nonumber \\
&= &  \frac{\pi}{4}\left[\left<E\right>-m_N\left<N_B-N_{\bar B}\right>\right] .
\end{eqnarray}

The above equation relates the transverse energy measured from the data and that
estimated from the thermal model. 
In the limit of large beam energies one has
\begin{eqnarray}
\lim_{\sqrt{s_{NN}}\rightarrow\infty}{\left<E_T\right>\over N_{\textrm{ch}}} 
&=&{\left<E_T\right>\over 0.6 N_{\textrm{decay}}},\nonumber\\
&=&{\pi\over 4}{1\over 0.6}{E\over 1.7 N} ,\nonumber\\
&=&0.77{E\over N},\nonumber\\
&\approx& 0.83~\textrm{GeV} . 
\end{eqnarray}
This value is close to the value measured at RHIC.
It should be noted that
the measured $E_T$
will be affected by the transverse collective flow 
and by the difference between chemical freeze-out and kinetic freeze-out 
temperatures and therefore the  description presented here
is only a qualitative one. 
An analysis including flow 
was presented in Fig. 17 of the review article by 
Kolb and Heinz~\cite{heinz} who show that this improves 
the agreement with the data at SPS and RHIC beam energies.
A detailed comparison  in the framework of a 
specific model with a single freeze-out temperature, 
has been made in Ref.~\cite{prorok}.

At higher 
energies, when $\mu_B$ nearly goes to 
zero, the transverse energy production is mainly due to the meson content in the matter.
The intersection points of lines of constant $E_T/N_{\textrm{ch}}$ and 
the freeze-out line give the values of $E_T/N_{\textrm{ch}}$ at the chemical freeze-out. 
Hence at freeze-out, given the values of $E_T/N_{\textrm{ch}}$ from the experimental 
measurements we can determine $T$ and $\mu_B$ of the system.

\begin{figure}
\begin{center}
\resizebox{0.75\columnwidth}{!}{%
\includegraphics{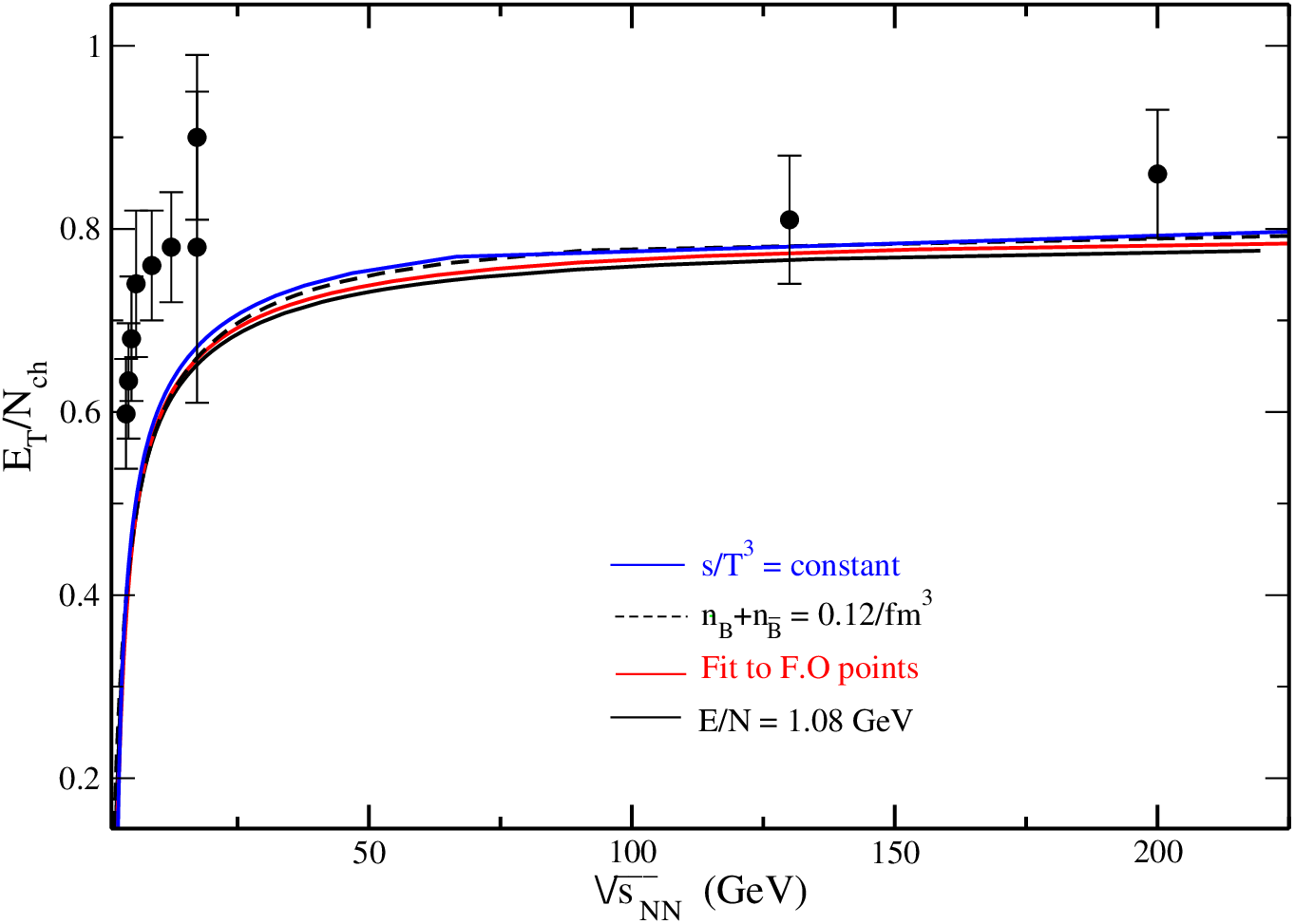}}
\caption{Comparison between experimental data for
 $E_T/N_{\textrm{ch}}$ with $\sqrt{s_{NN}}$ and the thermal model
using  $E/N = 1.08$ GeV as  well as other  freeze-out 
conditions~\cite{PBMS,Majiec,TawfikEntropy,tawfik2}.}
\label{etNchCM}
\end{center}
\end{figure}

For the most central collisions, the variation of $E_T/N_{\textrm{ch}}$ with center of mass energy
is shown in Fig.~\ref{etNchCM}. The data have been taken from 
Ref.~\cite{fopi,e802,e814_6GeV,wa98_17GeV,na49_17GeV,na49,star200GeV,phenix130GeV,phenixSyst},
and are compared with the corresponding calculation
from the thermal model with chemical freeze-out.
We have checked explicitly
that other freeze-out criteria discussed in the literature give almost identical results for the behavior of 
$E_T/N_{\textrm{ch}}$ as a function of $\sqrt{s_{NN}}$;
this is the case for the fixed baryon plus anti-baryon density
condition~\cite{PBMS}
and also for fixed normalized entropy density condition,
$s/T^3$ = 7~\cite{Majiec,TawfikEntropy,tawfik2}.
We have checked explicitly that the 
centrality behavior is well reproduced by 
the thermal hadronic resonance gas model.
\cite{raghu}.
%

In conclusion, we have discussed the connection between  $E_T/N_{\textrm{ch}}$  
and the ratio of primordial energy to primordial particle 
multiplicity, $E/N$,  from the thermal model.  
This
model, when  combined with  chemical freeze-out criteria 
explains the data over all available measurements for the $\sqrt{s_{NN}}$ 
behavior of $E_T/N_{\textrm{ch}}$. 
It has to be noted that  variables like
$E_T/N_{\textrm{ch}}$, the chemical freeze-out temperature $T_{\textrm{ch}}$, 
$N_{\textrm{decays}}/N_{\textrm{primordial}}$ 
and $N_{\textrm{ch}}/N_{\textrm{decays}}$ discussed
in this paper, show  saturation starting at SPS 
and continuing to higher center of mass energies. 
This observation along with the centrality independence of
$E_T/N_{\textrm{ch}}$ is not inconsistent with the simultaneity
of chemical and kinetic freeze-out at higher energies~\cite{sFO}.

\section*{Acknowledgement}
Three of us (JC,~RS,~DKS) would like to acknowledge the financial support of
the South Africa-India Science and Technology agreement.

\end{document}